\documentclass[aps,amsmath,amssymb,superscriptaddress, twocolumn]{revtex4}
\usepackage{graphicx}
\usepackage{dcolumn}
\usepackage{bm}

\usepackage{colordvi}
\usepackage{color}

\begin{document}

\title{Low-temperature photoluminescence of oxide-covered single-layer MoS$_2$}

\author{G. Plechinger}
\affiliation{Institut f\"ur Experimentelle und Angewandte Physik,
Universit\"at Regensburg, D-93040 Regensburg, Germany}
\author{F.-X. Schrettenbrunner}
\affiliation{Institut f\"ur Experimentelle und Angewandte Physik,
Universit\"at Regensburg, D-93040 Regensburg, Germany}
\author{J. Eroms}
\affiliation{Institut f\"ur Experimentelle und Angewandte Physik,
Universit\"at Regensburg, D-93040 Regensburg, Germany}
\author{D. Weiss}
\affiliation{Institut f\"ur Experimentelle und Angewandte Physik,
Universit\"at Regensburg, D-93040 Regensburg, Germany}
\author{C.\ Sch\"uller}
\affiliation{Institut f\"ur Experimentelle und Angewandte Physik,
Universit\"at Regensburg, D-93040 Regensburg, Germany}
\author{T.\ Korn}
\email{tobias.korn@physik.uni-regensburg.de}
\affiliation{Institut
f\"ur Experimentelle und Angewandte Physik, Universit\"at
Regensburg, D-93040 Regensburg, Germany}
\date{\today}

\begin{abstract}
  We present a photoluminescence study of single-layer MoS$_2$ flakes on  SiO$_2$ surfaces. We demonstrate that the luminescence peak position of flakes prepared from natural MoS$_2$, which varies by up to 25~meV between individual as-prepared flakes, can be homogenized by annealing in vacuum, which removes adsorbates from the surface. We use HfO$_2$ and Al$_2$O$_3$ layers prepared by atomic layer deposition to cover some of our flakes. We clearly observe a suppression of the low-energy luminescence peak observed for as-prepared flakes at low temperatures, indicating that this peak originates from excitons bound to surface adsorbates. We also observe different temperature-induced shifts of the luminescence peaks for the oxide-covered flakes. This effect stems from the different thermal expansion coefficients of the oxide layers and the MoS$_2$ flakes. It indicates that the single-layer MoS$_2$ flakes strongly adhere to the oxide layers and are therefore strained.
\end{abstract}

\maketitle

The dichalcogenide MoS$_2$ has a layered crystal structure, in which adjacent layers are only weakly coupled by Van der Waals forces, while intralayer coupling is due to covalent bonds. Therefore, flakes containing very few or even a single MoS$_2$ layer may easily be prepared on top of SiO$_2$ using the mechanical cleavage method that is well-established for graphene~\cite{Novoselov26072005}. While bulk MoS$_2$ is an indirect-gap semiconductor, single-layer MoS$_2$ flakes were recently shown to be strong emitters of photoluminescence~\cite{Heinz_PRL10,Splen_Nano10}, indicating a drastic change of the band structure to a direct gap, which was predicted by density functional theory calculations~\cite{Eriksson09}. Recent time-resolved measurements demonstrated short photocarrier lifetimes in few- and single-layer MoS$_2$~\cite{Korn11,Ruzicka11}. This makes single-layer MoS$_2$ flakes highly interesting for potential electro-optic devices, e.g., fast photodetectors.
\begin{figure*}[htb]%
\centering
\includegraphics*[width= \textwidth]{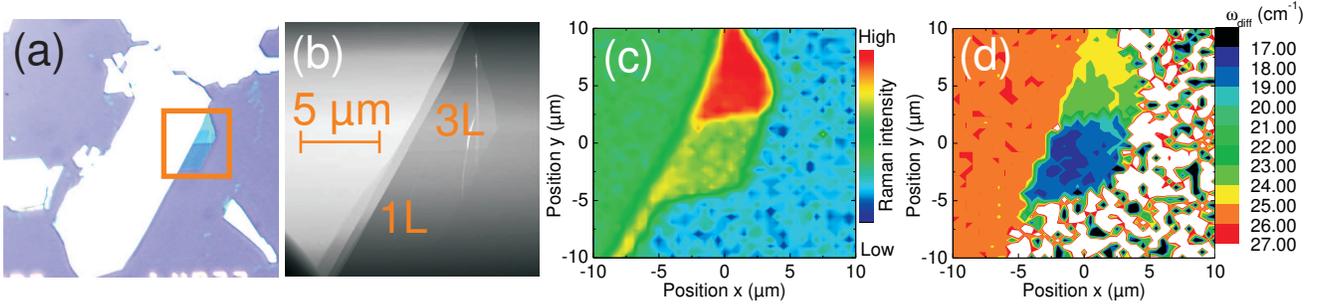}
\caption{(a) Optical micrograph of a MoS$_2$ flake on a SiO$_2$/Si wafer. The orange square indicates the area of the AFM scan shown in (b). (c) False color image showing the intensity of the Raman E$^1_{2g}$ mode. The scan area is similar to the AFM scan in (b). (d) False color image showing the peak frequency difference $\Delta\omega$ of the Raman A$^1_{g}$ and E$^1_{2g}$ modes. The scan area is identical to (c).}
\label{Fig1_charact.eps}
\end{figure*}
For  applications, such as  transistors~\cite{Kis_NatNano10}, the MoS$_2$  must be protected from the environment by a large-bandgap dielectric layer, such as Al$_2$O$_3$ or HfO$_2$.

Here, we present scanning Raman and temperature-dependent photoluminescence (PL) measurements of single-layer MoS$_2$ flakes, some of which have been covered by oxide layers using atomic layer deposition (ALD).
Our samples are prepared from natural MoS$_2$, and deposited on a Silicon wafer covered with a 300~nm thick SiO$_2$ layer (thermal oxide) using the transparent tape liftoff technique. For ease of orientation on the wafer, metal marker structures are lithographically defined prior to deposition of MoS$_2$. First characterization of the deposited flakes is performed with an optical microscope, and few-layer flakes are readily identifiable because of the phase shift induced by the thin flakes~\cite{castellanos-gomez:213116,Kis11} on top of the SiO$_2$ layer. Selected flakes are analyzed in more detail using atomic force microscopy (AFM), PL and scanning Raman measurements. For deposition of oxide layers, we first identify wafer pieces containing single-layer flakes, then cover the whole wafer piece with about 15~nm HfO$_2$ or Al$_2$O$_3$ using ALD. The samples were annealed prior to ALD at 450$^\circ$~C in a 10~percent hydrogen/90~percent nitrogen atmosphere, then immediately transferred to the ALD chamber.  Both dielectrics were grown at 250$^\circ$~C with alternating pulses of trimethylaluminum (TMA) and water for the Al$_2$O$_3$ layer and Tetrakis(dimethylamino)hafnium (TDMAH) and water for the HfO$_2$ layer, respectively. Scanning Raman spectroscopy measurements are performed at room temperature, details of the Raman setup  are published elsewhere~\cite{Korn11}. For micro-PL measurements, the samples are mounted in a He-flow cryostat. A microscope objective is used to focus the excitation laser (continuous-wave excitation at 532~nm) onto the sample and to collect the emitted PL, which is coupled into a spectrometer and detected using a CCD camera. The laser spot is adjusted such that only the single-layer part of the investigated flakes is illuminated.

First, we discuss the characterization of the MoS$_2$ flakes. Figure \ref{Fig1_charact.eps} (a) shows an optical micrograph of a MoS$_2$ flake:  thicker areas of the flake are opaque with a grey-white colour, thinner areas are distinguishable by different shades of blue, while the SiO$_2$ layer has a uniform, purple color. A small region of the flake depicted in Fig. \ref{Fig1_charact.eps} (a) is imaged using AFM (Fig. \ref{Fig1_charact.eps} (b)). Areas of uniform height are clearly seen, and the   step heights extracted from  AFM traces allow us to identify the areas marked '1L' and '3L' as single and triple layers of MoS$_2$. Scanning Raman measurements were performed on the same flake. Figure \ref{Fig1_charact.eps} (c) depicts a false color image in which the intensity of the MoS$_2$ E$^1_{2g}$ mode (in-plane optical vibration of Mo and S atoms) is color-coded. The Raman intensity maps the flake shape, with a large intensity for the intermediate-thickness area on the top part of the flake. Further analysis is possible due to the thickness dependence of the MoS$_2$ Raman modes~\cite{Heinz_ACSNano10}: while the A$^1_{g}$ mode increases in frequency with the flake thickness, the E$^1_{2g}$ anomalously softens due to increased dielectric screening~\cite{Molina11}, so that the difference of the mode frequencies is characteristic for a certain thickness. Figure \ref{Fig1_charact.eps} (d) shows this frequency difference, $\Delta\omega$, as a function of position, the image area is the same as in Fig. \ref{Fig1_charact.eps} (c). The lower part of the flake has $\Delta\omega=18$~cm$^{-1}$ (1 layer), the upper part shows a larger $\Delta\omega$ of 23~cm$^{-1}$~(3 layers), while the large, thick portion of the flake on the left has $\Delta\omega=25$~cm$^{-1}$ as expected for bulk-like flakes. Subsequent PL measurements confirm that strong PL is only observed on the single-layer part of the flake.

\begin{figure}[h]%
\centering
\includegraphics[width=0.9 \linewidth]{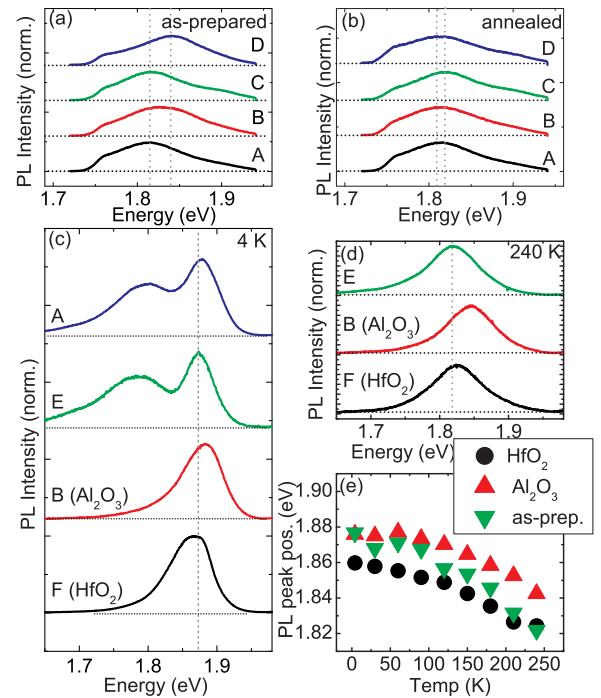}
\caption{(a) Room-temperature PL spectra of 4 different MoS$_2$ flakes after preparation.  (b) Room-temperature PL spectra for the same flakes shown in (a), after annealing the flakes in  vacuum at 150$^\circ$~C for 30 minutes. The dashed lines in both graphs indicate the range of the  PL peak positions of the flakes. (c) PL spectra of 4 different MoS$_2$ flakes measured at 4~K. Samples A and E are not covered by oxide layers.  (d)  PL spectra for 3 of the flakes shown in (a) measured at 240~K. The dashed lines in both graphs indicate the high-energy PL peak positions of the flakes. (e) PL peak position for the 3 flakes shown in (d) as a function of temperature.}
\label{Fig2_5panel_PL.eps}
\end{figure}

Next, we discuss the PL of  single-layer flakes at room temperature. Figure \ref{Fig2_5panel_PL.eps} (a) shows the room-temperature PL spectra for 4 different MoS$_2$ flakes prepared using the same natural MoS$_2$ source. We note that the PL peak position varies significantly from flake to flake, with a standard deviation of 10~meV and an average value of $E_{P}= 1821$~meV. After this first series of PL measurements, the samples were mounted in a cryostat and annealed in vacuum  at a temperature of 150$^\circ$~C for 30 minutes. Subsequent PL measurements of the same flakes, performed without breaking the vacuum, (Fig. \ref{Fig2_5panel_PL.eps}(b)) demonstrate that the PL peak position becomes more uniform, with a standard deviation of 1.5~meV and an average value of $E_{P}= 1815$~meV.
We interpret these observations as follows: since the single-layer flakes at the substrate surface are exposed to ambient conditions, water vapor and other adsorbates are likely to form on top of the flakes, leading to a change of the dielectric environment (as compared to air/vacuum) on top of the flakes, which in turn influences the exciton binding energy due to screening of the electron-hole Coulomb interaction. Further spectral shift can arise from the Stark effect, which can be caused by charged impurities on the flake surface. The PL peak position therefore shifts, and the magnitude of the shift depends on the coverage of the individual flake. These adsorbates are partially removed from the flake surface by the annealing process. The slight differences in the PL line shape between individual flakes, even after annealing, indicate inhomogeneous line broadening due to some remaining adsorbates.

Finally, we discuss the low-temperature PL experiments.
Figure~\ref{Fig2_5panel_PL.eps} (c) shows PL spectra of two as-prepared samples and samples covered with either Al$_2$O$_3$ or HfO$_2$, measured at 4~K. We clearly observe that for both as-prepared samples, the PL spectra show two distinct peaks, a broad, low-energy peak, and a more narrow high-energy peak. By contrast, the two oxide-covered flakes only show the high-energy peak, with a weakly pronounced low-energy shoulder. We may therefore infer that the low-energy peak observed in the as-prepared flakes stems from excitons bound to surface impurities, and that the oxide layers effectively suppress the formation of such impurity-bound excitons. This indicates that the annealing process and the heating of the samples in the ALD chamber effectively remove the majority of the surface impurities. We note that the peak positions for the two oxide-covered samples are shifted to higher (Al$_2$O$_3$) or lower (HfO$_2$) energy as compared to the as-prepared flakes. If dielectric screening of the  Coulomb interaction were the cause of the observed peak shifts, we would expect a blueshift of the PL, which should be larger for HfO$_2$ ($\epsilon_r \approx 20$) than for Al$_2$O$_3$ ($\epsilon_r \approx 10$). A series of temperature-dependent PL measurements demonstrates that the temperature-induced PL-peak shift is also very different for flakes with and without oxide layers (Fig.~\ref{Fig2_5panel_PL.eps} (e)): the largest shift in the temperature range from 4~K to 240~K is observed for the as-prepared flakes ($\Delta E =55$~meV), while HfO$_2$-($\Delta E =36$~meV) and Al$_2$O$_3$-covered ($\Delta E =34$~meV) flakes show significantly smaller shifts, so that at 240~K (Fig.~\ref{Fig2_5panel_PL.eps} (d)), both oxide-covered flakes are blue-shifted with respect to the as-prepared flakes. We may understand these observations as follows: the ALD deposition of the oxide layers occurs at 250$^\circ$~C, and during the subsequent cooling of the samples to room temperature and below, the MoS$_2$ flakes are strained due to the different thermal expansion coefficients of the MoS$_2$ flake and the top and bottom oxide layers, resulting in strain-induced changes of the band gap. Since the as-prepared flakes show a larger shift of the bandgap with T than the oxide-covered flakes, we may infer that the thermal expansion coefficient for MoS$_2$ in the temperature range between 240~K and 4~K is larger than that of the stacked Al$_2$O$_3$/HfO$_2$-SiO$_2$ sytem, resulting in \emph{tensile} strain during cooling. We note that the room-temperature thermal expansion coefficent of thermal oxide SiO$_2$,$\alpha_{SiO_2}=0.24\times 10^{-6}/K$~\cite{Lai05} is significantly lower than that of either of the two ALD-deposited oxides~\cite{Fei1995}  and  probably dominates the thermal expansion of the stacked system. The strain-induced bandgap shift also indicates that the MoS$_2$ flakes strongly adhere to the oxide if they are \emph{sandwiched} between top and bottom layers.

In conclusion, we have investigated single-layer MoS$_2$ by photoluminescence as a function of temperature for as-prepared, annealed and oxide-covered flakes. We observe that the PL peak position for different as-prepared flakes can be homogenized by annealing. In low-temperature measurements, we observe two PL peaks for the as-prepared flakes, while the lower-energy peak is suppressed for the flakes covered with oxide layers. The as-prepared flakes exhibit a stronger temperature-dependent band gap shift than the oxide-covered flakes, indicating that they are strained due to the different thermal expansion coefficients of the oxides and the MoS$_2$.
Financial support by the DFG via SFB689 and GRK1570 is gratefully acknowledged.

\end{document}